# Irreversible Thermodynamic of a New Model of The Collision Term of the Boltzmann Kinetic Equation Dealing with Gas Mixture Affected by a Centrifugal Field.

Taha Zakaraia Abdel Wahid [1, 2]

[1] Department of Basic Sciences, El-Gezeera High Institute for Engineering and Technology, Elmokatam City, Cairo, Egypt.

[2] Mathematic Department, Faculty of Science, Menofia University, Shebin El-Kom 32511, Egypt

Email: taha_zakaraia@yahoo.com

## Abstract

In the present paper, a new model of the collision term, of the Boltzmann Kinetic equation dealing with binary gas mixture affected by a centrifugal field, is introduced. The new model is representing a modification of the BGK (Bhatnagar-Gross-Krook) model of the Boltzmann kinetic equation to be suitable for studying the influence of centrifugal field affected on gases mixture. The scientific achievement of the new model is that it is a very simple model without any mathematical complexity; but without any loss of generality. We shed light upon that the new model is satisfied the conservation of energy, conservation of mass, and conservation of momentum laws, the second law of thermodynamic, and the principle of maximum entropy production that is like the BGK model itself. Complete mathematical model for a centrifugal field, used in Uranium enrichment process affected on gases mixture between two coaxial circular porous rotating cylinders, is presented which is an important application for our new model. We also, introduced the complete irreversible thermodynamics properties of the system for the first time at all, for the best of our knowledge. The entropy, entropy flux, entropy production, thermodynamic forces, kinetic coefficients are presented for our model system. The ratios between the different contributions of the internal energy changes are predicted via the extended Gibbs equation. The main important implementation of the model is the uranium enrichment that is used in nuclear energy generation or nuclear weapons, many medical applications and various industrial amazing applications. The flow of a binary gases mixture of $UF_6$, and $N_2$ between rotating cylinders is the first important suggested problem enforcement of the new model.

**Introduction:**

Uranium enrichment increases the proportion of the fissile isotope U-**235** about five- or six-fold from the 0.7% of U-**235** found in natural uranium. Uranium enrichment is a physical process, usually relying on the small mass difference between atoms of the two isotopes U-**238** and U-**235** [1]. We should shed light up on that, one gram of U-235 could release enough energy during fission to raise the temperature of 66 million gallons of water from 25$^o$ C to 100$^o$ C while, to accomplish the same sort of feat by burning pure octane; it would require 1.65 million gallons of the fuel.

The uranium enrichment processes in commercial use today require the uranium to be in a gaseous form and hence use the compound uranium hexafluoride ($UF_6$). This becomes a gas at only 56$^o$ C under atmospheric pressure, but is readily contained in steel cylinders as a liquid or solid under pressure [1]. Gas is supplied and centrifugal force tends to compress it in the outer region, but thermal agitation tends to redistribute the gas molecules throughout the whole volume. Light molecules are favored in this effect, and their concentration is higher near the center axis. By various means, a countercurrent flow of $UF_6$ gas is established that tends to carry the heavy and light isotopes to opposite ends of the rotor [1, 2]. The American physicist J. W. Beams et al are the first who successfully demonstrate the centrifugal process of isotope separation by separating two chlorine isotopes through a vacuum ultracentrifuge [3].

Centrifugal separations may be classified as either analytical or preparative. In 'analytical centrifugation', the objective is to monitor particle sedimentation behavior in order to characterize particle properties, e.g., molecular weight, shape, and association. Today, centrifuges are routinely used in a variety of disciplines including the medical, pharmaceutical, mineral, chemical, dairy, food, and agricultural industries. Available centrifuge designs and configurations seem almost as numerous as the applications themselves [4].

**Mathematical and physical problem:**

Consider the unsteady flow of a gas mixture between two infinite coaxial circular cylinders with radii $r = r_1$ and $r_2$ such that is $(r_1 < r_2)$. The outer cylinder is at rest, while the inner cylinder rotates at a constant angular velocity $\Omega$. The rotation of the inner cylinder cases a centrifugal force $(m\Omega^2 r)$ where $m$ is the mass of gas atom. An identical constant temperature $T_w$ is maintained at the surfaces of the cylinders with complete momentum and energy accommodation and diffuse reflection of incident gas molecules with equilibrium distribution functions $f_1$ and $f_2$ at the same temperature $T_w$.

We introduce a cylindrical coordinate system $(r,\theta, z)$, where the $Oz$ axis coincides with the axis of the cylinders, $r$ is the distance from this axis, and $\theta$ is an azimuthally coordinate. The ax-symmetric $z$-independent state of a gas mixture is determined by the molecular-velocity distribution function $f(r,c_z,c_r,c_\theta,t)$, where $c_r, c_\theta$ and $c_z$ are the orthogonal components of the molecular velocity in the radial, azimuthal, and axial directions, respectively [5]. In the velocity space, we also introduce a cylindrical coordinate system with an axis parallel to the $Oz$ axis. Let $c_n$ denotes the velocity component that is lying in a plane perpendicular to the cylinders' axis and $\psi$ be the angle between this component and the radial direction from the axis of symmetry. The orthogonal velocity components $c_r$ and $c_\theta$, and the polar coordinates of the velocity are related by the formulas Fig.(1): $c_r = c_n \sin\psi, c_\theta = c_n \cos\psi$

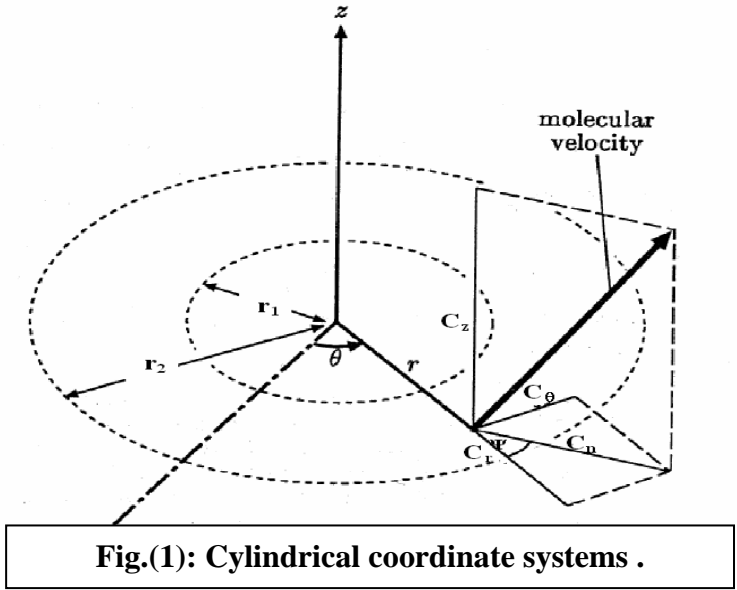

**Fig.(1): Cylindrical coordinate systems .**

The gas moves along azimuthal direction, while its radial velocity is zero [5]. We assumed that the distribution functions satisfied the kinetic equation with the new modified BGK collision term. The kinetic equation for both gases is written as [6-13]:

(1) $$\frac{\partial f_A}{\partial t} + \vec{c}\cdot\frac{\partial f_A}{\partial \vec{r}} + \frac{\vec{F}_A}{m_A}\cdot\frac{\partial f_A}{\partial \vec{c}} = \nu_{AA}(f_{A0}-f_A) + \nu_{AB}(f_{B0}-f_A)$$

(2) $$\frac{\partial f_B}{\partial t} + \vec{c} \cdot \frac{\partial f_B}{\partial \vec{r}} + \frac{\vec{F}_B}{m_B} \cdot \frac{\partial f_B}{\partial \vec{c}} = \nu_{BB}(f_{B0} - f_B) + \nu_{BA}(f_{A0} - f_B)$$

where $\nu_{AA}, \nu_{AB}, \nu_{BB}$ and $\nu_{BA}$ are (gas A- gas A, gas A - gas B, gas B-gas B and gas B-gas A) new modified collision frequencies respectively that are given by us in the guide of the references [12-17]:

(3)
$$\nu_{AA} = \frac{n_0}{2}\left[\pi\left(\frac{d_A + d_A}{2}\right)^2\right]\left[\sqrt{\frac{8KT_w}{\pi m_A}}\right]\left[Exp\left(\frac{-m_A \Omega^2 r^2}{2KT_w}\right)\right]$$

,

$$\nu_{AB} = n_0\left[\pi\left(\frac{d_A + d_B}{2}\right)^2\right]\left[\sqrt{\frac{8KT_w}{\pi m_A}}\right]\left[Exp\left(\frac{-m_A \Omega^2 r^2}{2KT_w}\right)\right]\left[\frac{\frac{m_A m_B}{m_A + m_B}}{m_B}\right],$$

$$\nu_{BB} = \frac{n_0}{2}\left[\pi\left(\frac{d_B + d_B}{2}\right)^2\right]\left[\sqrt{\frac{8KT_w}{\pi m_B}}\right]\left[Exp\left(\frac{-m_B \Omega^2 r^2}{2KT_w}\right)\right]$$

and $\nu_{BA} = n_0\left[\pi\left(\frac{d_B + d_A}{2}\right)^2\right]\left[\sqrt{\frac{8KT_w}{\pi m_B}}\right]\left[Exp\left(\frac{-m_B \Omega^2 r^2}{2KT_w}\right)\right]\left[\frac{\frac{m_A m_B}{m_A + m_B}}{m_A}\right]$

where $n_0$ is the gas concentration at rest, $f_{\beta 0}$ are the local Maxwellian distribution function denoted by:

(4) $$f_{\beta 0} = n_0(2\pi RT_w)^{-\frac{3}{2}}\left(1 + \frac{c_r u_{r\beta}}{RT_w} + \frac{c_\theta u_{\theta\beta}}{RT_w}\right)\exp\left(\frac{-c^2}{2RT_w}\right),$$

$\beta = A$ or $B$

As the BGK model of the Boltzmann kinetic equation satisfied the conservation of energy, conservation of mass, conservation of momentum laws, the second law of thermodynamic, and the principle of maximum entropy production. Then our new model, also, satisfied the conservation of energy, conservation of mass, conservation of momentum laws, the second law of thermodynamic, and the principle of maximum entropy production because our new model had the same dependent the equilibrium and nonequilibrium distribution functions. Let us write the solution of Equation (1, 2) as suggested by Kashmarov [5] in the form:

$$(5)\, f_\beta = \begin{cases} f_{1\beta} = n_0 (2\pi R T_w)^{-\frac{3}{2}} \left(1 + \dfrac{c_r u_{r1\beta}}{RT_w} + \dfrac{c_\theta u_{\theta 1\beta}}{RT_w}\right) \\ \exp\left(\dfrac{-c^2}{2RT_w}\right) : \alpha \leq \psi \leq \pi - \alpha \\ f_{2\beta} = n_0 (2\pi R T_w)^{-\frac{3}{2}} \left(1 + \dfrac{c_r u_{r2\beta}}{RT_w} + \dfrac{c_\theta u_{\theta 2\beta}}{RT_w}\right) \\ \exp\left(\dfrac{-c^2}{2RT_w}\right) : \pi - \alpha \leq \psi \leq 2\pi + \alpha \end{cases}$$

where $u_{ri\beta}$ and $u_{\theta i\beta}$ are eight unknown functions of time $t$ and the distance variable $r$ where $\beta = A$ or $B$ and $i = 1$ or $2$.

Using Grad's moment method [18- 20] multiplying equations (7) by $Q_j(\vec{c})$ and integrating over all values of $\vec{c}$, we obtain the transfer equations in the forms:

for gas A:

$$(6)\quad \frac{\partial}{\partial t}\left(\int Q_i f_A \underline{dc}\right) + \frac{1}{r}\frac{\partial}{\partial r}\left(r \int Q_i c_r f_A \underline{dc}\right) - \frac{1}{r}\int c_\theta^2 f_A \frac{\partial Q_i}{\partial c_r}\underline{dc} +$$
$$\frac{1}{r}\int c_r c_\theta f_A \frac{\partial Q_i}{\partial c_\theta}\underline{dc} - \left(\int (\Omega r^2) f_A \frac{\partial Q_i}{\partial c_r}\underline{dc}\right) =$$
$$v_{AA}\int Q_i (f_{A0} - f_A)\underline{dc} + v_{AB}\int Q_i (f_{B0} - f_A)\underline{dc}$$

and for gas B:

$$(7)\quad \frac{\partial}{\partial t}\left(\int Q_i f_B \underline{dc}\right) + \frac{1}{r}\frac{\partial}{\partial r}\left(r \int Q_i c_r f_B \underline{dc}\right) - \frac{1}{r}\int c_\theta^2 f_B \frac{\partial Q_i}{\partial c_r}\underline{dc} +$$
$$\frac{1}{r}\int c_r c_\theta f_B \frac{\partial Q_i}{\partial c_\theta}\underline{dc} - \left(\int (\Omega r^2) f_B \frac{\partial Q_i}{\partial c_r}\underline{dc}\right) =$$
$$v_{BB}\int Q_i (f_{B0} - f_B)\underline{dc} + v_{BA}\int Q_i (f_{A0} - f_B)\underline{dc}$$

where $Q_i$ is a function of the velocity. The moment $\overline{Q_i}$ of $Q_j(\vec{c})$ is given by the integrals over the velocity distance from the relation [18- 20],

$$(8)\quad \overline{Q_i} = \int Q_i(\vec{c}) f_\beta \underline{dc} = \int_\alpha^{\pi-\alpha}\int_0^\infty \int_{-\infty}^\infty Q_i f_{1\beta} c_n dc_z dc_n d\psi +$$
$$\int_{\pi-\alpha}^{2\pi+\alpha}\int_0^\infty \int_{-\infty}^\infty Q_i f_{2\beta} c_n dc_z dc_n d\psi$$

where $Q_i = Q_i(\vec{c}), i = 1, 2$ and $\underline{dc} = c_n dc_z dc_n d\psi$

## 3- The Non-Equilibrium Thermodynamic Properties of The System:

The everyday resorts to the linear theory of the thermodynamics of irreversible processes still constitute great interests [18-24]. This is associated both with the general theoretical importance of this theory and its numerous applications in various branches of science. It is unquestionable that the concept of entropy has played an essential role both in the physical and biological sciences [25-29]. Thus, we start the thermodynamic investigations of the problem from the evaluation of the entropy $S$ per unit mass of the gas, which is written as [23-29]:

$$(9)\quad \bar{S} = -\int f \ln f\, d\bar{C} = -\left(\int_{-\infty}^{\infty}\int_{-\infty}^{0}\int_{-\infty}^{\infty} f_1 \ln f_1\, \overline{dC} + \int_{-\infty}^{\infty}\int_{0}^{\infty}\int_{-\infty}^{\infty} f_2 \ln f_2\, \overline{dC}\right)$$

The entropy flux vector

$$(10)\quad \vec{J} = -\left(\int_{-\infty}^{\infty}\int_{-\infty}^{0}\int_{-\infty}^{\infty} \vec{C} f_1 \ln f_1 \overline{dC} + \int_{-\infty}^{\infty}\int_{0}^{\infty}\int_{-\infty}^{\infty} \vec{C} f_2 \ln f_2 \overline{dC}\right)$$

Accordingly, the entropy production has the form [23-29]:

$$(11)\quad \sigma = \frac{dS}{dt} = \frac{\partial S}{\partial t} + \vec{\nabla}\cdot \vec{J} \geq 0$$

The generalized thermodynamic forces are as follows [20-23]:

$$(12)\quad X_1 = \frac{\Delta r}{n}\nabla_r n,\quad X_2 = \frac{\Delta r}{U}\nabla_r U,\quad X_3 = M^*$$

where $M^* = \frac{\Omega r}{\sqrt{2RT}}$ is the centrifugal Mach number and $\Delta r$ is the thickness of the layer adjacent to the inner cylinder in units of the mean free path, the distance between two collisions of the gas particles, in dimensionless form.

After calculating the thermodynamic forces and the entropy production, we can obtain the kinetic coefficients $L_{ij}$ from the relationship between the entropy production and the thermodynamic forces, via the form [20, 23]:

$$(13).\quad \sigma_S(y,t) = \sum_i \sum_j L_{ij} X_i X_j = (X_1\ \ X_2\ \ X_3)\begin{pmatrix} L_{11} & L_{12} & L_{13} \\ L_{21} & L_{22} & L_{23} \\ L_{31} & L_{32} & L_{33} \end{pmatrix}\begin{pmatrix} X_1 \\ X_2 \\ X_3 \end{pmatrix} \geq 0$$

This constitute the restriction on the sign of phenomenological coefficients $L_{ij}$, which arise because of the second law of the thermodynamics [20, 23-29], which can be deduced from the standard results in algebra. The necessary and sufficient conditions for $\sigma_s(y,t) \geq 0$ are fulfilled by the determinant

$$(14)\quad |L_{ij} + L_{ji}| \geq 0,$$

and all its principal minors are non-negative too. Another restriction on $L_{ij}$ was established by Onsager (1931). He found that, besides the restriction on the sign, the phenomenological coefficients verify important symmetry properties. Invoking the principle of microscopic reversibility and using the theory of fluctuations, Onsager was able to demonstrate the symmetry property denoted by,

(15) $L_{ij} = L_{ji}$ ,

which is called the Onsager's reciprocal relations.

The Gibbs formula for the variation of the internal energy applied to the system, $dU(y,t)$ is as follows:

(16) $dU(y,t) = dU_S(y,t) + dU_V(y,t)$ .

The internal energy change due to the variation of the extensive variables, such as entropy $dU_S$, volume $dU_V$, are respectively read for a gas [23], as follows:

(16) $dU = dU_S + dU_V$

where $dU_S = T_0 dS$ and $dU_V = P_0 dV$ . Here $dV = \frac{-dn}{n^2}$ , $P_0 = n_0 T_0$ , $dS = \frac{\partial S}{\partial t}\delta t + \frac{\partial S}{\partial r}\delta r + \frac{\partial S}{\partial \theta}\delta \theta , dn = \frac{\partial n}{\partial r}\delta r$ .

Thus, we had finished the wholly presented of the new model and its thermodynamics prediction.

**Conclusion:**

We had introduced a new model of the collision term for the Boltzmann kinetic equation that is suitable for binary gas mixture suffered from strong centrifugal field in the uranium enrichment processes, which is also suitable for many various applications in industries. We also, introduced the complete irreversible properties of the system for the first time at all, for the best of our knowledge.

# References

[1] Raymond L. Murray, Keith E. Holbert "Nuclear Energy: An Introduction to the Concepts, Systems, and Applications of Nuclear Processes" (Seventh Edition), Elsevier, USA, 2015, pp 243-248. ( ISBN: 978-0-12-416654-7)

[2] Cutler J. Cleveland and Christopher G. Morris, Handbook of Energy. Volume II: Chronologies, Top Ten Lists, and Word Clouds, Elsevier, USA, 2014. ( ISBN : 978-0-12-417013-1 )

[3] D.N. Taulbee, A. Furst "Centrifugation" ,Reference Module in Chemistry, Molecular Sciences and Chemical Engineering, from Encyclopedia of Analytical Science(Second Edition), 2005, pp 469-481.

[4] M. M. R. Williams, Mathematical Methods in Particle Transport Theory, Butterworth, London, 1971. ISBN: 9780408700696)

[5] Taha Zakaraia Abdel Wahid " The Effect of Lorentz and Centrifugal Forces on Gases and Plasma." , LAMBERT Academic Publishing, Germany, ISBN: 978-620-2-05504-8, (2018).

[6] Taha Zakaraia Abdel Wahid" Travelling Waves Solution of the Unsteady Flow Problem of a Rarefied Non-Homogeneous Charged Gas Bounded by an Oscillating Plate." Mathematical Problems in Engineering 2013; 2013(ID 503729): 1-13.

[7] Taha Zakaraia Abdel Wahid " Kinetic and Thermodynamic Study of The Thermal Radiation Effect on The Gases." , LAMBERT Academic Publishing, Germany, ISBN: ISBN 978-613-9-92664-0, (2018).

[8] Taha Zakaraia Abdel Wahid "Kinetic and Thermodynamic Treatments of Unsteady Gaseous Plasma Flows", LAMBERT Academic Publishing, Germany, ISBN: 978-613-9-90736-6, (2017).

[9] Taha Zakaraia Abdel Wahid, Travelling Waves Solution of the Unsteady Problem of Binary Gas Mixture Affected by a Nonlinear Thermal Radiation Field, American Journal of Physics and Applications. Vol. 2, No. 6, 2014, pp. 121-134. doi: 10.11648/j.ajpa.20140206.13.

[10] A. M. Abourabia and T. Z. Abdel Wahid "Kinetic and Thermodynamic Treatment for the Rayleigh Flow Problem of an Inhomogeneous Charged Gas Mixture", J. Non-Equilib. Thermodyn. 37 (2012), 119–141.

[11] A. M. Abourabia and T. Z. Abdel Wahid "Kinetic and Thermodynamic Treatment of a Neutral Binary Gas Mixture Affected by a Nonlinear Thermal Radiation Field." Can. J. Phys. 90: 137–149 (2012).

[12] C. Cercignani, Theory and Application of the Boltzmann Equation, Elsevier, New York, 1975.

[13] C. Cercignani, Mathematical Methods in Kinetic Theory, 2$^{nd}$ edition, Plenum Press, New York, 1990.

[14] E. A. Johnson and P. J. Stopford; Arch. Mech.,Vol. 37, 1-2, pp 29-37, Warszawa, (1985).

[15] E. A. Johnson and P. J. Stopford, Physics of fluids, Vol. 27, p106, (1984).

[16] E. A. Johnson, United Kingdom Atomic Energy Authority Report AERER 10461 (Her Majesty's Stationery office, London),(1982).

[17] G. C. Pomraning ; Non-continuum shear flow in a centrifugal field ,G. E. T. I. series, report No: APED-4374, 63APE22, (1963).

[18] B. Chan Eu, "Kinetic Theory and Irreversible Thermodynamics", Wiley, NY, USA, 1992.

[19] D. Jou and J. Casas-Vázquez, G. Lebon, "Extended Irreversible Thermodynamic", Springer -Verlag, Berlin, Germany, (1993).

[20] Taha Zakaraia Abdel Wahid "Kinetic and Irreversible Thermodynamic study of Plasma and Neutral Gases.", LAMBERT Academic Publishing, Germany, (2014), (ISBN: 978-3-659-62296-0).

[21] V. Zhdanov and V. Roldughin, "Non-Equilibrium Thermodynamics and Kinetic Theory of Rarefied Gases" , Physics Uspekhi, 4 (4) (1998), 349-378.

[22] V. M. Zhdanov and V. I. Roldugin, "Nonequilibrium Thermodynamics of a Weakly Rarefied Gas Mixture" Zh. Eksp. Teor. Fiz., 109, (1996), 1267–1287.

[23] G. Lebon , D. Jou , J. Casas-Vàzquez "Understanding Non-equilibrium Thermodynamics : Foundations, Applications, Frontiers " Springer-Verlag Berlin , Heidelberg, (2008).

[24] F. Sharipov "Reciprocal relations based on the non-stationary Boltzmann equation", Physica A 391, 1972-1983 (2012).

[25] R. M. Velasco, L. S. García-Colín and F. J. Uribe " Entropy Production: Its Role in Non-Equilibrium Thermodynamics" Entropy, 2011, 13, 82-116; (doi:10.3390/e13010082)

[26] A. B. De Castro, "Continuum Thermomechanics" Birkhauser, Verlag, Basel, Switzerland, (2005).

**Author Contributions**

T. Z. W. wrote the manuscript, made the derivation, prepared figures and reviewed the manuscript.

**Additional Information**

**Competing financial interests:** The authors declare no competing interests.